\documentclass[reprint,amsmath,amssymb,aps]{revtex4-1}
\usepackage[colorlinks,citecolor=blue,linkcolor=blue,anchorcolor=blue,filecolor=blue,urlcolor=blue]{hyperref}
\usepackage{graphicx}
\usepackage{ulem}
\usepackage{dcolumn}
\usepackage{bm}
\usepackage{cancel}
\usepackage{numprint}

\begin{document}

\title{Dark matter effects in modified teleparallel gravity}

\author{S. G. Vilhena$^1$, M. Dutra$^{1,2}$, O. Louren\c{c}o$^{1,2}$, and P. J. Pompeia$^1$}
\affiliation{
\mbox{$^1$Departamento de F\'isica, Instituto Tecnol\'ogico de Aeron\'autica, DCTA, 12228-900, 
S\~ao Jos\'e dos Campos, SP, Brazil} \\ 
\mbox{$^2$Universit\'e de Lyon, Universit\'e Claude Bernard Lyon 1, CNRS/IN2P3, IP2I Lyon, UMR 5822, F-69622, Villeurbanne, France}
}

\date{\today}

\begin{abstract}

This work investigates dark matter (DM) effects in compact objects in modified teleparallel gravity (MTG) in which a modification of Teleparallel Equivalent to General Relativity is used. We applied a tetrad to the modified field equations where a set of relations is found. The conservation equation allows us to rewrite our Tolman-Oppenheimer-Volkoff equations with an effective gravitational coupling constant. As input to these new equations, we use a relativistic mean-field (RMF) model with dark matter content included, obtained from a Lagrangian density with both, hadronic and dark particle degrees of freedom, as well as the Higgs boson, used as a mediator in both sectors of the theory. Through numerical calculations, we analyze the mass-radius diagrams obtained from different parametrizations of the \mbox{RMF-DM} model, generated by assuming different values of the dark particle Fermi momentum and running the free parameter coming from the MTG. Our results show that it is possible for the system simultaneously support more DM content, and be compatible with recent astrophysical data provided by LIGO and Virgo Collaboration, as well as by NASA's Neutron star Interior Composition Explorer (NICER).
\end{abstract}

\maketitle

\section{Introduction\label{sec:intro}}

The problem of galaxy rotation curves is possibly one of the most iconic indications that General Relativity (GR) in the presence of ordinary matter is not the final theory of gravitation \cite{Rubin1980,Begeman1991,Corbelli2000}.
Indeed, in order to solve this (and other) problem(s), modifications to GR have become a trend and two main lines of research have emerged. 

One of them consists in modifying the theory of gravitation. In this branch, one can keep the underlying Riemann manifold and modify the gravitational action integral, for instance, by taking an action that is proportional to a function of the scalar curvature $(R)$ and/or the Ricci $(R_{\mu\nu})$ and Riemann $(R_{\mu\nu\rho\sigma})$ tensors: $f\left(R\right)$,$f\left(R,R_{\mu\nu},R_{\mu\nu\rho\sigma}\right)$, and so on \cite{Faraoni2010,Capozziello2011,Bajardi2022,Rinaldi2023,Nojiri2017,Shapiro2021,PJP2016}. In other proposals, the manifold is changed. Of particular interest here are the theories built on a Weitzenb\"ock manifold -- the so-called ``teleparallel theories'' \cite{Hayashi1979,Maluf2001,Pereira2004,PJP2003,Cai_2016,Ferraro2022,Blagojevic2020,PJP2021,Hohmann2022}. In this manifold, gravitation is manifest by means of torsion instead of curvature. Many theories can be built in the Weitzenb\"ock manifold, including one that is equivalent to GR, known as the ``Teleparallel Equivalent of General Relativity'' (TEGR). The gravitational action in this case is built as a specific combination of three quadratic invariants of the torsion tensor. Different combinations of these invariants lead to different theories where the equivalence with GR is no longer valid. The criticisms concerning modified theories of gravity usually claim the lack of consistency to describe simultaneously systems of stellar, galaxy, and cosmological scales \cite{Pardo2020} .

The second line of research explores the modification of the matter content in the context of GR. The models propose the existence of non-ordinary matter which does not interact with (ordinary) matter by means of the known interactions of the Standard Model of particles. Some candidates for dark matter may eventually interact weakly, like WIMPs~\cite{bertone,cand1}. For galaxy rotation curves, dark matter (DM) has been considered a strong candidate to solve this problem. The criticism concerning the existence of DM is the lack of evidence from the point of view of particle physics.

Usually, these two lines of research compete with each other and are considered in excluding scenarios. In other words, few proposals (as far as the authors are concerned) consider that a final solution to the problems presented by GR may be found in a ``joint venture'' with both modified gravity and DM. This is what shall be explored in this work, where the effects of both modified gravity and DM will be analyzed in a neutron star. This system is particularly interesting since this is an extremely dense object and the relativistic effects are very relevant.

The modified gravity model considered here is the same presented in Ref.~\cite{silas}, which is a teleparallel theory with arbitrary coefficients for the three quadratic invariants of the torsion tensor. For the solution of the modified Tolman-Oppenheimer-Volkoff equations generated from this theory, it is required the knowledge of suitable equations of state. For this purpose, we use here a realistic parametrization of a widely used hadronic model~\cite{rev3,dutra2014,silva} with dark matter included along with protons, neutrons, and leptons. We show how the teleparallel theory favors the increasing of the dark matter amount in the system, and verify that it is possible to generate neutron stars in agreement with very recent astrophysical observational data, such as those related to gravitational waves detection, performed by LIGO and Virgo Collaboration~\citep{Abbott_2017,Abbott_2018,Abbott_2020,Abbott_2020-2}, those furnished by the NASA's Neutron star Interior Composition Explorer (NICER) mission regarding the pulsars PSR~J0030+0451~\citep{Riley_2019,Miller_2019} and PSR~J0740+6620~\citep{Riley_2021,Miller_2021} (where we also adding the data extracted from \cite{Fonseca_2021} for the last one), and the PSR~J0952-0607~\cite{Romani_2022}, discovered by Bassa et al. in 2017~\citep{Bassa_2017}. We present all these findings in the following way: in Sec.~\ref{sec:mtg} we briefly revise the derivation of the teleparallel theory developed in Ref.~\cite{silas}. In Sec.~\ref{rmf-dm} we provide the main equations of state obtained from the relativistic mean-field model in which we also include interacting dark matter through the Higgs mechanism. In Sec.~\ref{sec:results} we show the mass-radius diagrams determined from the teleparallel theory and discuss the importance of this theory for the total DM content included in the system. We finalize the paper in Sec.~\ref{summ} by presenting a summary and our concluding remarks.

\section{Modified Teleparallel Gravity}
\label{sec:mtg}

Differently from GR where gravity is represented by the curvature of spacetime, Teleparallel Theories describe gravity solely by the spacetime torsion. Torsion is the anti-symmetric part of the connection and is represented by
\begin{equation}
\Gamma_{\mu\nu}^{\rho}-\Gamma_{\nu\mu}^{\rho}=T_{\hphantom{\mu}\mu\nu}^{\rho}\,,\label{Eq_TorsionDef}
\end{equation}
and in GR, it is null. The connection components in GR are given by the Christoffel symbols, which are completely determined by the metric tensor, $g_{\mu\nu}$, and its derivatives. In this sense, $g_{\mu\nu}$ plays the role of the fundamental field in GR and others theories that are constructed on the Riemannian manifold. In Teleparallel theories, the connection is called ``Weitzenb\"ock connection" and is completely determined by the tetrad field, $e_{\hphantom{\mu}\mu}^{a}$ and its derivatives -- in other words, the tetrad is the fundamental field. From a  geometrical point of view, the tetrad maps objects of the spacetime manifold in their corresponding counterparts of the tangent space; for instance, the spacetime metric and the tangent metric are mapped with the help of the tetrad by the contraction:
\begin{equation}
g_{\mu\nu}=\eta_{ab}e_{\hphantom{\mu}\mu}^{a}e_{\hphantom{\mu}\nu}^{b}\,,\label{Eq_MetricTetradRel}
\end{equation}
As mentioned previously, the Weitzenb\"ock connection is determined by the tetrad and its derivative: 
\begin{equation}
\Gamma_{\mu\nu}^{\rho}=e_{\hphantom{\nu}a}^{\rho}\partial_{\mu}e_{\hphantom{\nu}\nu}^{a}\,.\label{Eq_WeitConnectionDef}
\end{equation}

While in GR the Lagrangian is essentially the Ricci scalar, in Teleparallel Gravity the Lagrangian is built with a linear combination of quadratic invariants of the torsion tensor{[}Eq.~(\ref{Eq_TorsionDef}){]}. There are only three quadratic invariants that can be constructed with torsion. In our model we propose a general combination of these invariants \cite{silas}:
\begin{equation}
\mathcal{L}=-\frac{\beta_{1}}{4}T^{\rho\mu\nu}T_{\rho\mu\nu}-\frac{\beta_{2}}{2}T^{\rho\mu\nu}T_{\mu\rho\nu}+\beta_{3}T^{\rho}T_{\rho}\,,\label{Eq_MTGLagrangian}
\end{equation}
The difference between our Lagrangian and that from TEGR is the presence of three free parameters, namely, $\beta_{1},\beta_{2}\text{ and }\beta_{3}$ -- in TEGR, $\beta_{1}=\beta_{2}=\beta_{3}=1$. The variational principle for Eq. (\ref{Eq_MTGLagrangian}) leads us to a modified set of field equations (in comparison with TEGR), although the general structure is preserved:

\begin{equation}
\partial_{\rho}\left(4e\Sigma_{f}^{\hphantom{f}\lambda\rho}\right)+4e\Sigma_{d}^{\hphantom{d}\lambda\rho}T_{\hphantom{d}f\rho}^{d}-ee_{f}^{\lambda}\Sigma_{ijk}T^{ijk}=-2\chi ee_{f}^{\rho}T_{\hphantom{\lambda}\rho}^{\lambda}\,.\label{Eq_FieldEq}
\end{equation}
Above, $e=\det e_{\hphantom{a}\mu}^{a}$ and the super-potential $\Sigma^{\mu\nu\rho}$ carries the free parameters in its definition:
\begin{align}
\Sigma^{\mu\nu\rho}\equiv\frac{\beta_1}{4}T^{\mu\nu\rho}+&\frac{\beta_2}{4}\left(T^{\nu\mu\rho}-T^{\rho\mu\nu}\right)\nonumber\\
&+\frac{\beta_3}{2}\left(g^{\mu\rho}T^{\nu}-g^{\mu\nu}T^{\rho}\right)\,.   
\label{eq:NewSigma}
\end{align}
In this sense, TEGR is a special case of our Modified Teleparallel Gravity~(MTG). It is worth mentioning that MTG and New General Relativity~(NGR) are equivalents when our free parameters relate to those ones from NGR, as can be verified in~\cite{silas}.

 When we apply a tetrad for static spherical objects, 
\begin{equation}
e_{\hphantom{a}\mu}^{a}=\left(\begin{array}{cccc}
\gamma_{00} & 0 & 0 & 0\\
0 & \gamma_{11}\sin\theta\cos\phi & r\cos\theta\cos\phi & -r\sin\theta\sin\phi\\
0 & \gamma_{11}\sin\theta\sin\phi & r\cos\theta\sin\phi & r\sin\theta\cos\phi\\
0 & \gamma_{11}\cos\theta & -r\sin\theta & 0
\end{array}\right)\,,\label{Eq_Tetamu}
\end{equation}
we find a set of three independent equations which are compatible with the conservation equation only under a specific condition, namely:
\begin{equation}
2\beta_{3}-\beta_{1}-\beta_{2}=0\,.\label{Eq_Constraint}
\end{equation}
Eq. (\ref{Eq_Constraint}) is obviously respected by TEGR. This constraint is very useful to rewrite the set of equations {[}Cf. Eq.(9) in Ref.\cite{silas}{]} with a redefinition of the gravitational coupling constant as an effective one, $\frac{\chi}{\beta_{3}}=\bar{\chi}$. We obtain a set of equations that resemble the TOV equations of GR:
\begin{equation}
\begin{cases}
\bar{\chi}\gamma_{11}^{3}\rho r^{2}=2\gamma_{11}^{\prime}r+\gamma_{11}^{3}-\gamma_{11}\\
-\bar{\chi}\gamma_{00}\gamma_{11}^{2}p r^{2}=2\gamma_{00}^{\prime}r-\gamma_{00}\gamma_{11}^{2}+\gamma_{00}\\
p^{\prime}=-\left(p+\rho\right)\frac{1}{\gamma_{00}}\gamma_{00}^{\prime}
\end{cases}\,.\label{Eq_SetEq}
\end{equation}

The first and second equations of (\ref{Eq_SetEq}) are both dependent on the effective gravitational coupling constant. Besides, the first equation is dependent on the energy density while the second one is dependent on the pressure. The third equation depends on both the pressure and energy density and it is the covariant conservation equation of the energy-momentum tensor. As we see, the difference between MTG and TEGR lies in the free parameter $\beta_{3}$ dividing $\chi$ as stated above. As consequence, we do not see differences in the field equations between those of our model and those from GR in a vacuum. So, the form of the Schwarzschild solution remains as the one of GR, as well as the description of planetary motions, deflection of light rays, and other applications of this solution without matter -- the difference may appear in terms of boundary conditions only. The structure of the equations allows us to use similar boundary conditions on a star to those used in GR.

The first equation of Eq. (\ref{Eq_SetEq}) becomes
\begin{equation}
u^{\prime}=\frac{1}{2}\bar{\chi}\rho r^{2}\,,\label{Eq_ubarprime}
\end{equation}
when the following change of variable is introduced 
\begin{equation}
\frac{1}{\gamma_{11}^{2}}=1-\frac{2u\left(r\right)}{r}\,.\label{Eq_ChaVar}
\end{equation}
The combination of the last two equations of (\ref{Eq_SetEq}) and (\ref{Eq_ChaVar}) allows us to correlate pressure and energy density with the derivative of the pressure in presence of the effective $\bar{\chi}$:
\begin{equation}
p^{\prime}=-\frac{p+\rho\left(p\right)}{\left(r^{2}-2ur\right)}\left(\frac{1}{2}\bar{\chi}p r^{3}+u\right)\,.\label{Eq_pbarprime}
\end{equation}
As expected, Eqs. (\ref{Eq_ubarprime}) and (\ref{Eq_pbarprime}) are analogous to those of GR (TOV equations) with the effective gravitational coupling constant. In this sense, the mass distribution of a compact object is completely determined by both expressions, and the equations of state (EoS) used as input.

\section{Hadronic Models with Dark Matter Component}
\label{rmf-dm}

The Lagrangian density related to the hadronic part of the model used in MTG reads
\begin{align}
&\mathcal{L}_{\mbox{\tiny HAD}} = \overline{\psi}(i\gamma^\mu\partial_\mu - M_{\mbox{\tiny nuc}})\psi 
+ g_\sigma\sigma\overline{\psi}\psi 
- g_\omega\overline{\psi}\gamma^\mu\omega_\mu\psi
\nonumber \\ 
&- \frac{g_\rho}{2}\overline{\psi}\gamma^\mu\vec{b}_\mu\vec{\tau}\psi
+\frac{1}{2}(\partial^\mu \sigma \partial_\mu \sigma - m^2_\sigma\sigma^2)
- \frac{A}{3}\sigma^3 - \frac{B}{4}\sigma^4 
\nonumber\\
&-\frac{1}{4}F^{\mu\nu}F_{\mu\nu} 
+ \frac{1}{2}m^2_\omega\omega_\mu\omega^\mu 
-\frac{1}{4}\vec{B}^{\mu\nu}\vec{B}_{\mu\nu} 
+ \frac{1}{2}m^2_\rho\vec{b}_\mu\vec{b}^\mu,
\label{dlag}
\end{align}
with the nucleon and exchanged mesons ($\sigma$, $\omega$, and $\rho$) fields given by $\psi$, $\sigma$, $\omega^\mu$, and $\vec{b}_\mu$, respectively, with masses denoted by $M_{\mbox{\tiny nuc}}$, $m_\sigma$, $m_\omega$, and $m_\rho$. The free parameters of the model are the coupling constants $g_\sigma$, $g_\omega$, $g_\rho$, $A$, and $B$. Finally, the tensors presented in Eq.~\eqref{dlag} are $F_{\mu\nu}=\partial_\mu\omega_\nu-\partial_\nu\omega_\mu$ and $\vec{B}_{\mu\nu}=\partial_\mu\vec{b}_\nu-\partial_\nu\vec{b}_\mu$. We address the reader to Refs.~\cite{rev3,dutra2014,silva} for more details regarding such kind of relativistic mean-field (RMF) models applied to symmetric and asymmetric nuclear matter. 

Concerning DM, we remark that the possibility of coupling this component with ordinary matter comes from the Higgs sector of the theory, in which one considers, for the present case, DM particles and nucleons simultaneously exchanging Higgs bosons. This procedure has been performed recently as one can see, for instance, in Refs.~\cite{rmfdm13,rmfdm1,rmfdm2,rmfdm3,abdul,rmfdm6,rmfdm11,rmfdm10,rmfdm8,rmfdm7,rmfdm12,dmnosso1,dmnosso2,dmnosso3,laura-tolos,sagun}. In this case, the entire system, namely, hadrons and DM, is described by the following Lagrangian density
\begin{align}
\mathcal{L} &= \overline{\chi}(i\gamma^\mu\partial_\mu - M_\chi)\chi
+ \xi h\overline{\chi}\chi +\frac{1}{2}(\partial^\mu h \partial_\mu h - m^2_h h^2)
\nonumber\\
&+ f\frac{M_{\mbox{\tiny nuc}}}{v}h\overline{\psi}\psi + \mathcal{L}_{\mbox{\tiny HAD}},
\label{dlagtotal}
\end{align}
with the Dirac field $\chi$ representing the dark fermion of mass $M_\chi=200$~GeV (lightest neutralino), and the scalar field $h$ denoting the Higgs boson with mass $m_h=125$~GeV. The strength of the Higgs-nucleon interaction is regulated by the quantity $fM_{\mbox{\tiny nuc}}/v$, with $v=246$~GeV being the Higgs vacuum expectation value. The Higgs-dark particle coupling is controlled by the constant $\xi$. The field equations for the model are obtained by the use of the mean-field approximation~\cite{rev3,dutra2014}, which consists in taking $\sigma\rightarrow \left<\sigma\right>\equiv\sigma$, $\omega_\mu\rightarrow \left<\omega_\mu\right>\equiv\omega_0$, $\vec{b}_\mu\rightarrow \left<\vec{b}_\mu\right>\equiv b_{0(3)}$, $h\rightarrow \left<h\right>\equiv h$. The final forms of these equations are given as follows,
\begin{align}
m^2_\sigma\,\sigma &= g_\sigma n_s - A\sigma^2 - B\sigma^3 
\\
m_\omega^2\,\omega_0 &= g_\omega n, 
\\
m_\rho^2\,b_{0(3)} &= \frac{g_\rho}{2}n_3, 
\\
[\gamma^\mu (&i\partial_\mu - g_\omega\omega_0 - g_\rho b_{0(3)}\tau_3/2) - M^*]\psi = 0,
\\
m^2_h\,h &= \xi n_s^{\mbox{\tiny DM}} + f\frac{M_{\mbox{\tiny nuc}}}{v}n_s
\label{hfield}
\\
(\gamma^\mu &i\partial_\mu - M_\chi^*)\chi = 0,
\end{align}
with $\tau_3=1$ for protons and $-1$ for neutrons, and effective dark particle and nucleon masses written as
\begin{eqnarray}
M^*_\chi = M_\chi - \xi h,
\end{eqnarray}
and
\begin{eqnarray}
M^* = M_{\mbox{\tiny nuc}} - g_\sigma\sigma - f\frac{M_{\mbox{\tiny nuc}}}{v}h,
\end{eqnarray}
respectively. The densities, or the sources of the fields, equivalently, are $n_s={n_s}_p+{n_s}_n$, $n=n_p + n_n$,
$n_3=\left<\overline{\psi}\gamma^0{\tau}_3\psi\right> = n_p - n_n=(2y_p-1)n$, 
\begin{eqnarray}
n_s^{\mbox{\tiny DM}} = \left<\overline{\chi}\chi\right> = 
\frac{\gamma M^*_\chi}{2\pi^2}\int_0^{k_F^{\mbox{\tiny DM}}} \hspace{-0.5cm}\frac{k^2dk}{(k^2+M^{*2}_\chi)^{1/2}},
\end{eqnarray}
and
\begin{align}
n_{s_{p,n}} = \left<\overline{\psi}_{p,n}\psi_{p,n}\right> =
\frac{\gamma M^*}{2\pi^2}\int_0^{k_{F_{p,n}}} \hspace{-0.5cm}\frac{k^2dk}{(k^2+M^{*2})^{1/2}}.
\end{align}
The symbols $p,n$ stand for protons and neutrons, and $\gamma=2$ is the degeneracy factor. The proton fraction is $y_p=n_p/n$, and the nucleon densities are $n_{p,n}=\gamma{k_F^3}_{p,n}/(6\pi^2)$, with ${k_F}_{p,n}$ and $k_F^{\mbox{\tiny DM}}$ being, respectively, the Fermi momenta related to protons/neutrons and to the dark particle. With regard to this last quantity, we keep it as a free parameter in our analysis performed in the next Section.

By using Eq.~\eqref{dlagtotal} it is possible to construct the main equations of state of the hadronic system with DM content, namely, energy density and pressure, both determined from the energy-momentum tensor associated to~$\mathcal{L}$, $T^{\mu\nu}$,  as $\epsilon=\left<T_{00}\right>$ and $P=\left<T_{ii}\right>/3$. They are explicitly written as
\begin{align} 
\epsilon &= \frac{m_{\sigma}^{2} \sigma^{2}}{2} +\frac{A\sigma^{3}}{3} +\frac{B\sigma^{4}}{4} 
-\frac{m_{\omega}^{2} \omega_{0}^{2}}{2} - \frac{m_{\rho}^{2} b_{0(3)}^{2}}{2} 
\nonumber\\
&+ g_{\omega} \omega_{0} \rho + \frac{g_{\rho}}{2} 
b_{0(3)} n_3 + \frac{m_h^2h^2}{2} + \epsilon_{\mathrm{kin}}^{p} + \epsilon_{\mathrm{kin}}^{n} + \epsilon_{\mathrm{kin}}^{\mbox{\tiny DM}},
\label{eden}
\end{align}
and
\begin{align}
P &= -\frac{m_{\sigma}^{2} \sigma^{2}}{2} - \frac{A\sigma^{3}}{3} - \frac{B\sigma^{4}}{4} + \frac{m_{\omega}^{2} \omega_{0}^{2}}{2} + \frac{m_{\rho}^{2} b_{0(3)}^{2}}{2}
\nonumber\\
& - \frac{m_h^2h^2}{2} + P_{\mathrm{kin}}^{p} + P_{\mathrm{kin}}^{n} + P_{\mathrm{kin}}^{\mbox{\tiny DM}},
\label{press}
\end{align}
with respective kinetic contributions,
\begin{eqnarray}
\epsilon_{\mbox{\tiny kin}}^{\mbox{\tiny DM}} &=& \frac{\gamma}{2\pi^2}\int_0^{k_F^{\mbox{\tiny DM}}}\hspace{-0.3cm}k^2(k^2+M^{*2}_\chi)^{1/2}dk,
\label{ekindm}
\\
P_{\mbox{\tiny kin}}^{\mbox{\tiny DM}} &=& 
\frac{\gamma}{6\pi^2}\int_0^{{k_F^{\mbox{\tiny DM}}}}\hspace{-0.5cm}\frac{k^4dk}{(k^2+M^{*2}_\chi)^{1/2}},
\label{pkindm}
\\
\epsilon_{\mbox{\tiny kin}}^{p,n} &=& \frac{\gamma}{2\pi^2} \int_0^{{k_F}_{p,n}}k^2({k^{2}+M^{*2}})^{1/2}dk,\,\,\,\mbox{and}
\\
P_{\mbox{\tiny kin}}^{p,n} &=&  
\frac{\gamma}{6\pi^2} \int_0^{k_{F\,{n,p}}}\hspace{-0.5cm}  
\frac{k^4dk}{\left({k^{2}+M^{*2}}\right)^{1/2}}.
\end{eqnarray}

 We use the NL3* parametrization, defined in Ref.~\cite{snl3}, for the hadronic part of the system. This parameter set was studied recently along with more than 400 other ones and has been verified as a model compatible with neutron stars, and finite nuclei properties~\cite{brett-jerome}. For the latter, it was analyzed in Ref.~\cite{brett-jerome} data related to ground state binding energies, charge radii, and giant monopole resonances of a set of spherical nuclei, namely, $^{16}\rm O$, $^{34}\rm Si$, $^{40}\rm Ca$, $^{48}\rm Ca$, $^{52}\rm Ca$, $^{54}\rm Ca$, $^{48}\rm Ni$, $^{56}\rm Ni$, $^{78}\rm Ni$, $^{90}\rm Zr$, $^{100}\rm Sn$, $^{132}\rm Sn$, and $^{208}\rm Pb$.

\section{Neutron stars properties (mass-radius profiles)}
\label{sec:results}

In Sec.~\ref{sec:mtg}, we demonstrated that MTG provides a set of equations totally equivalent to those from TEGR when the free parameter, $\beta_{3}$, is equal to unity. As stated above, different values of $\beta_{3}$ give us a renormalized gravitational coupling constant, as presented in the teleparallel equations~(TPE), namely, Eqs.~(\ref{Eq_ubarprime}) and~(\ref{Eq_pbarprime}). In this sense, in our model, this free parameter plays a central role, and it is important to verify how it modifies the mass-radius profiles when dark matter is considered in the system. In order to obtain such profiles, it is necessary to solve the TPE and for this aim, we furnish as input the total energy density and the total pressure as input, both written as
\begin{align}
\rho = \epsilon + \frac{\mu_e^4}{4\pi^2}
+ \frac{1}{\pi^2}\int_0^{\sqrt{\mu_\mu^2-m^2_\mu}}\hspace{-0.6cm}dk\,k^2(k^2+m_\mu^2)^{1/2},
\end{align}
and
\begin{align}
p = P + \frac{\mu_e^4}{12\pi^2}
+\frac{1}{3\pi^2}\int_0^{\sqrt{\mu_\mu^2-m^2_\mu}}\hspace{-0.5cm}\frac{dk\,k^4}{(k^2+m_\mu^2)^{1/2}},
\end{align}
with the two last terms of each equation representing, respectively, massless electrons of chemical potential $\mu_e$, and muons with mass $m_\mu=105.7$~MeV and chemical potential $\mu_\mu$. Besides that, the system is submitted to charge neutrality and chemical equilibrium, conditions represented by 
\begin{align}
n_p - n_e  = n_\mu,
\label{charge}
\end{align}
and
\begin{align} 
\mu_n - \mu_p = \mu_e,
\label{mueq}
\end{align}
where $\mu_e=(3\pi^2n_e)^{1/3}$, \mbox{$n_\mu=[(\mu_\mu^2 - m_\mu^2)^{3/2}]/(3\pi^2)$}, and $\mu_\mu=\mu_e$. Electrons and muons densities are denoted by $n_e$ and $n_\mu$. The solution of the TPE is constrained to $p(r=0)=p_c$ (central pressure) and $u(r=0) = 0$. On the other hand, at the star surface, we have $p(r=R) = 0$ and $u(r=R)\equiv M$, with $R$ defining the radius of the star. With regard to the neutron star crust, we use for this region an equation of state named BPS and provided by Baym, Pethick, and Sutherland~\cite{bps}, in a density range of $0.158\times10^{-10}\,\mbox{fm}^{-3} \leqslant n\leqslant 
0.891\times10^{-2}\,\mbox{fm}^{-3}$.

\begin{figure}[!htb]
\centering
\includegraphics[scale=0.33]{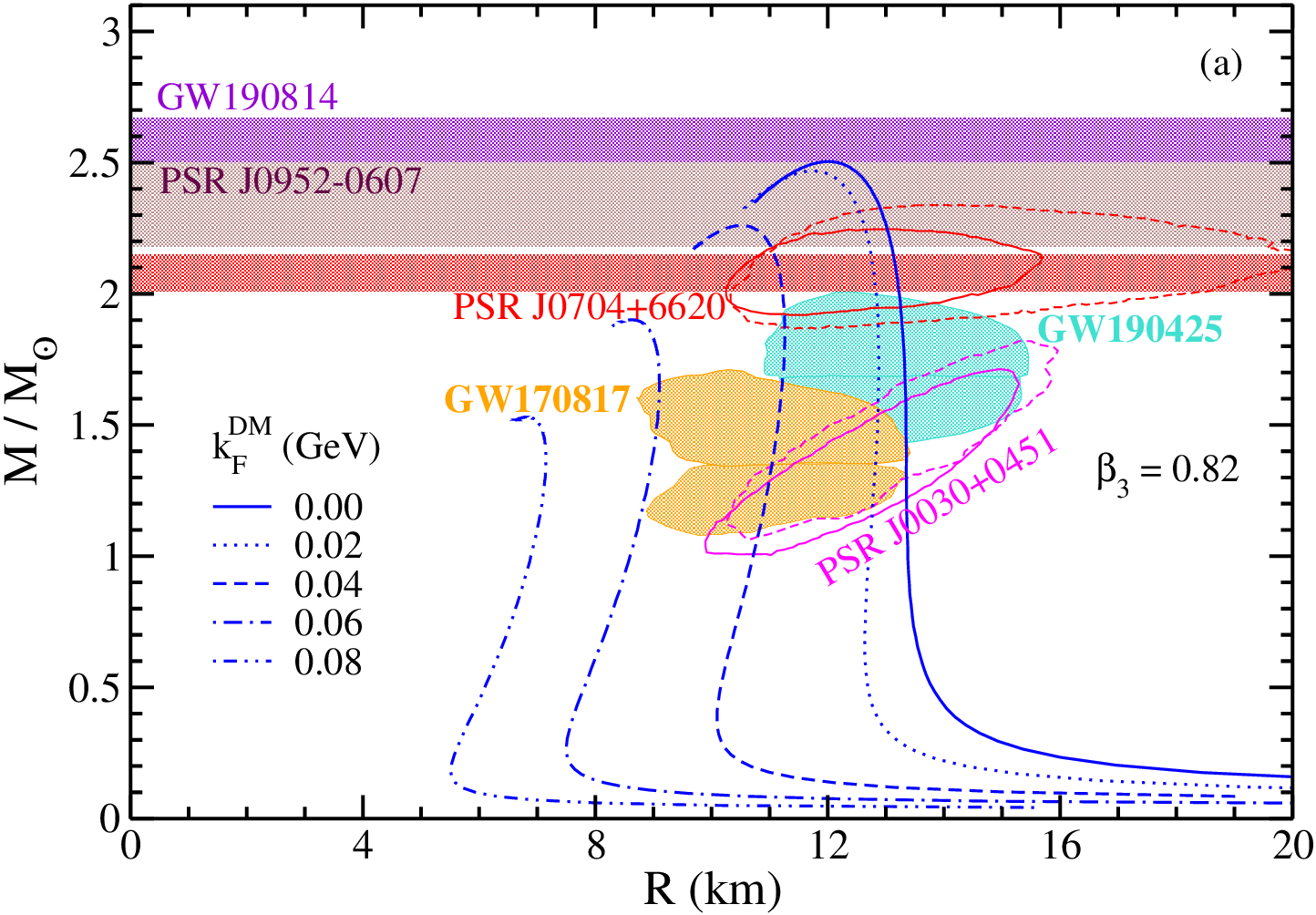}
\includegraphics[scale=0.33]{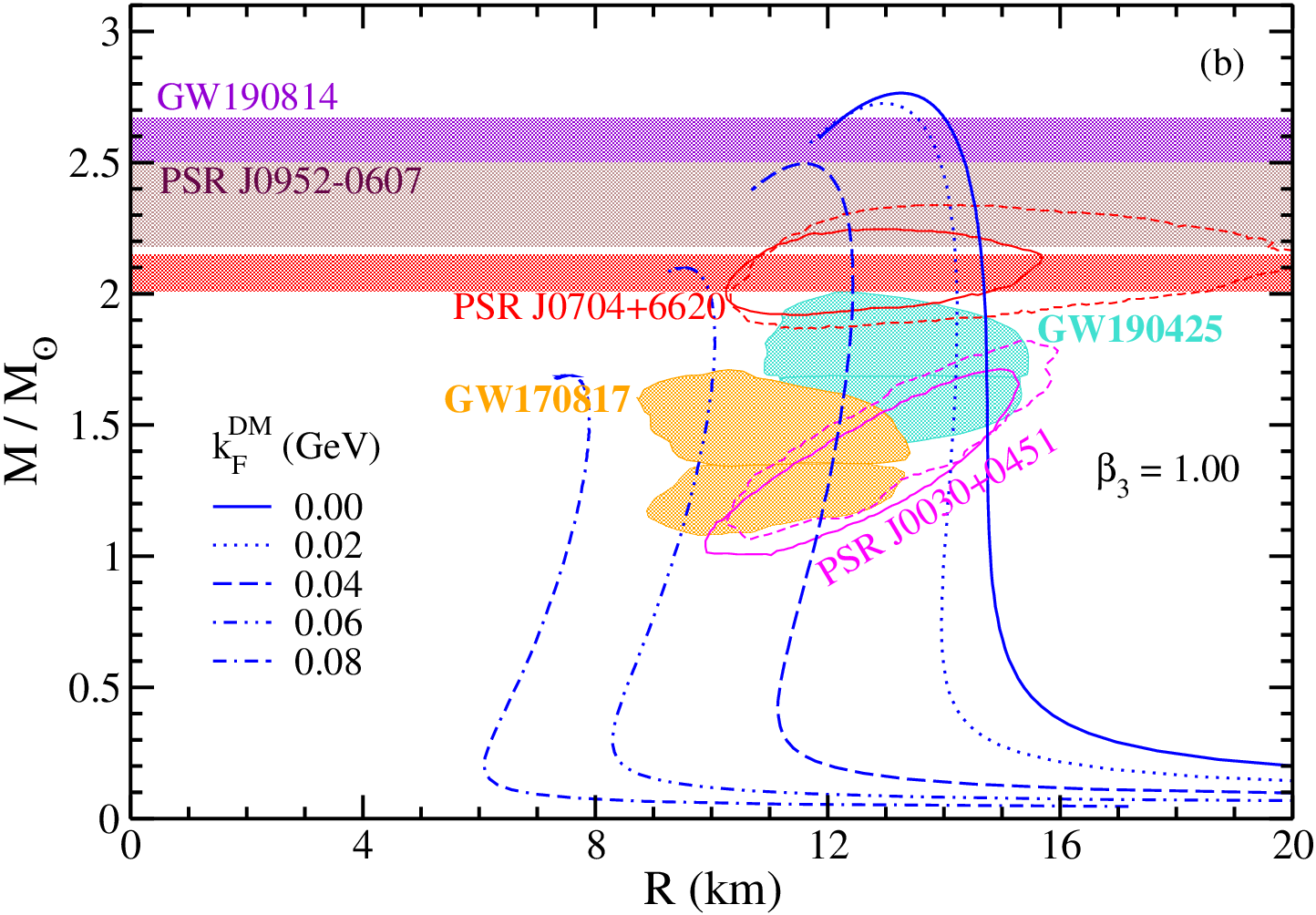}
\includegraphics[scale=0.33]{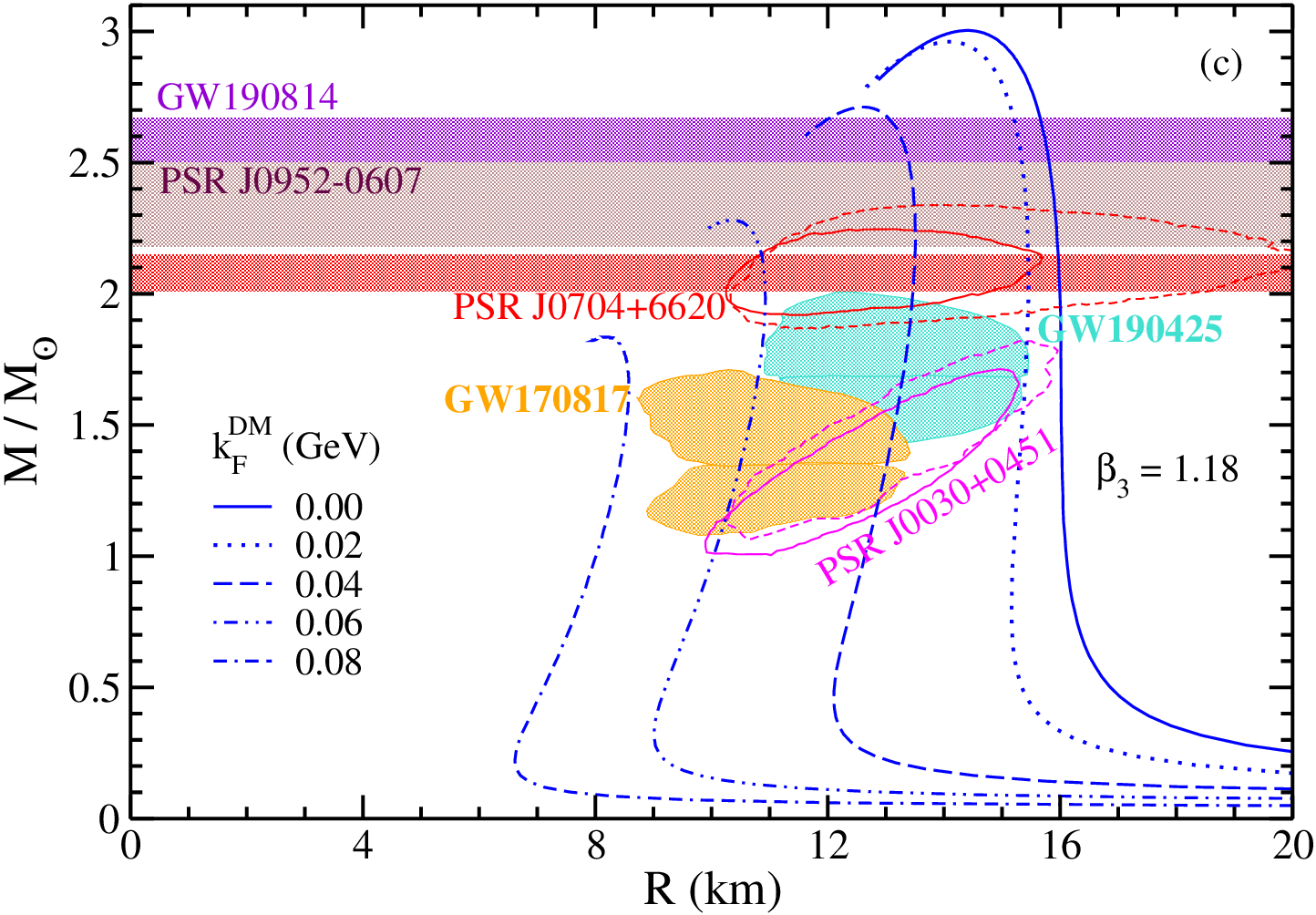}
\caption{Mass-radius diagrams constructed from the NL3* parametrization for different values of $k_F^{\mbox{\tiny DM}}$ and $\beta_3$, namely, (a) $\beta_3 = 0.82$, (b) $\beta_3 = 1.00$, and (c) $\beta_3 = 1.18$. The contours are related to data from the NICER mission, namely, PSR~J0030+0451~\citep{Riley_2019,Miller_2019} and PSR~J0740+6620~\citep{Riley_2021,Miller_2021}, the GW170817 event~\citep{Abbott_2017,Abbott_2018}, and the GW190425 event~\cite{Abbott_2020-2}, all of them at $90\%$ credible level. The red (brown) horizontal lines are related to PSR~J0740+6620~\citep{Fonseca_2021} (PSR~J0952+0607~\citep{Romani_2022}). The recent observational constraint on neutron star mass, GW190814~\citep{Abbott_2020}, is shown as the violet horizontal lines.}
\label{Fig1}
\end{figure}

\begin{figure*}[!htb]
\centering
\includegraphics[scale=0.33]{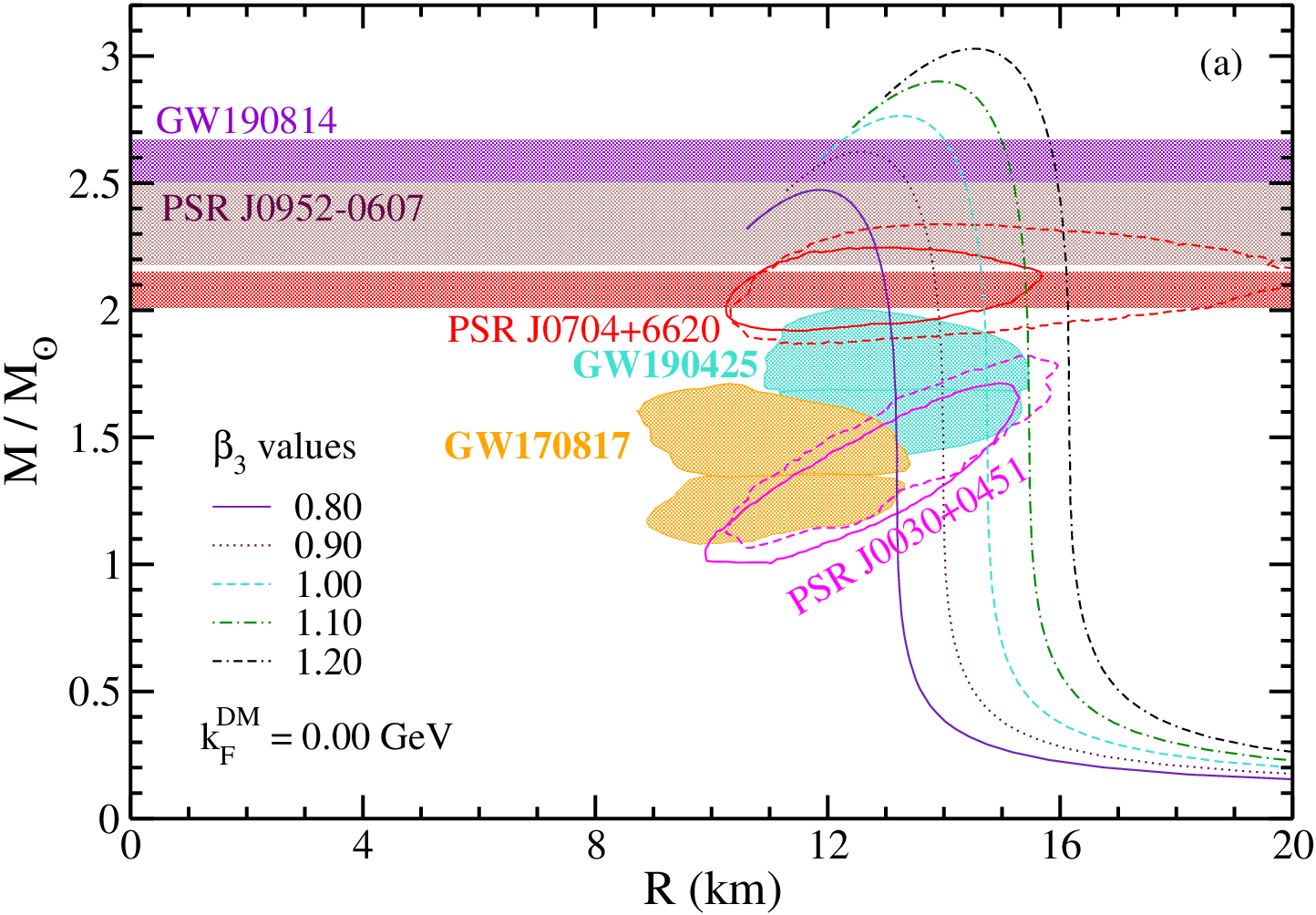}
\includegraphics[scale=0.33]{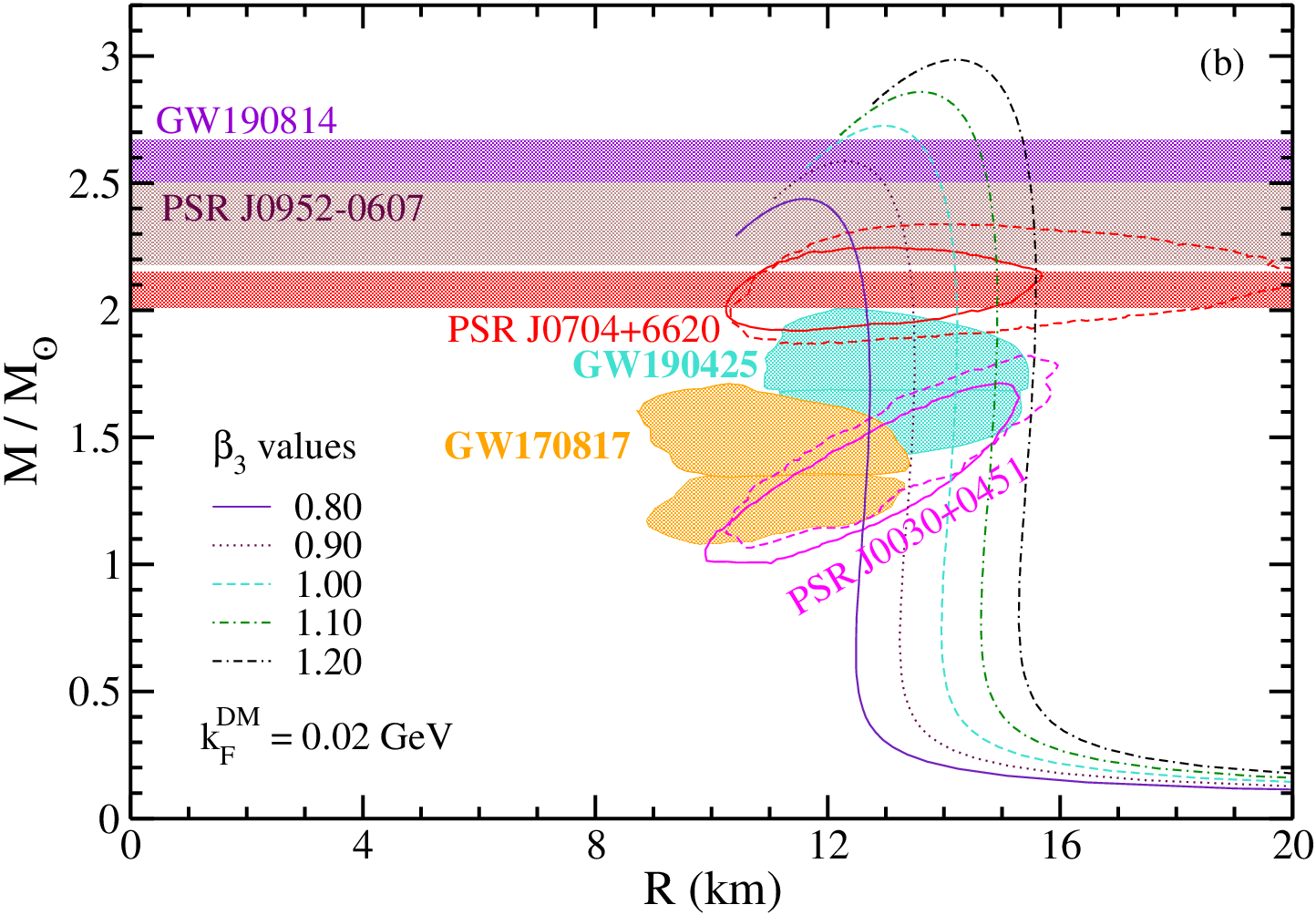}
\includegraphics[scale=0.33]{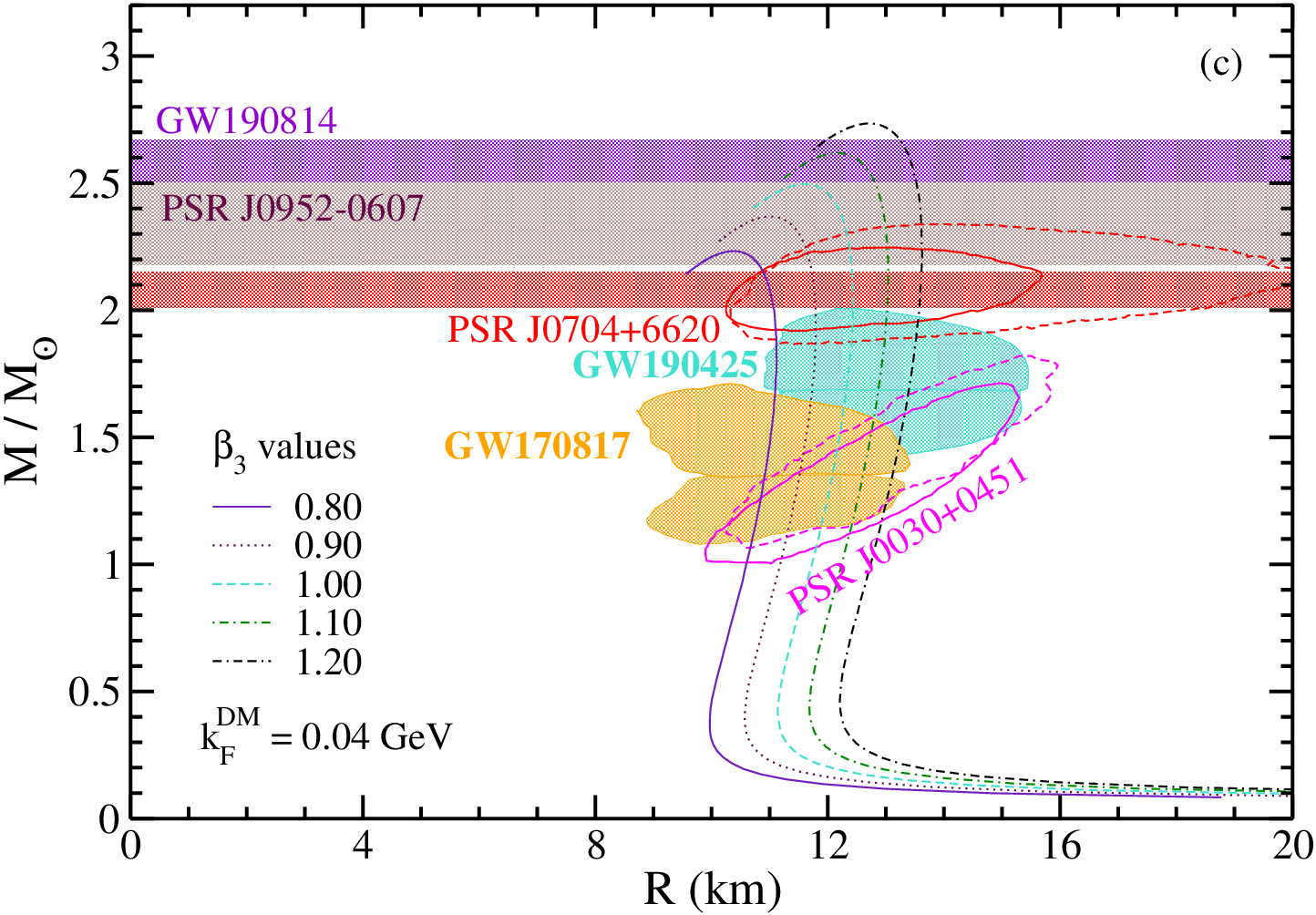}
\includegraphics[scale=0.33]{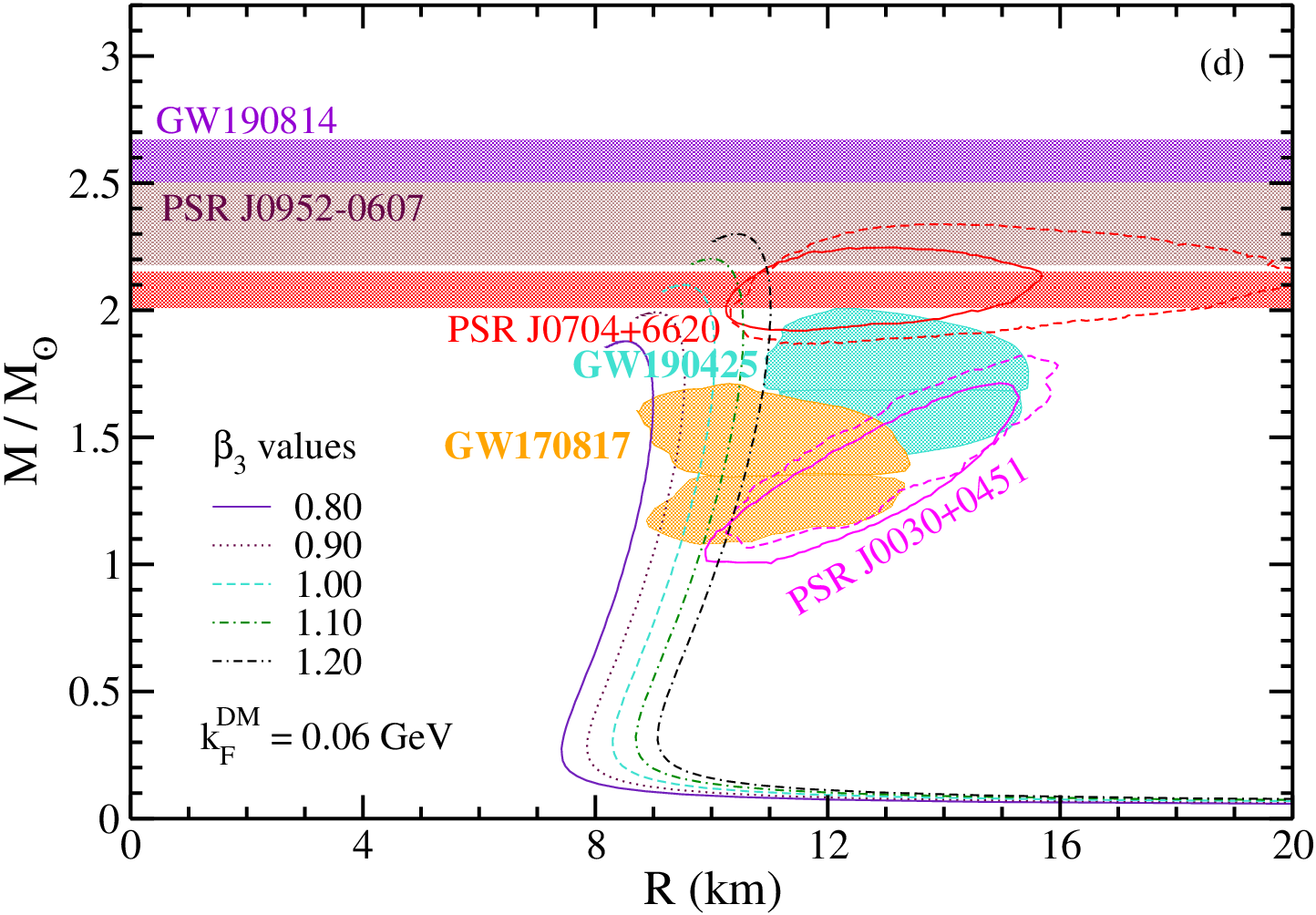}
\caption{Mass-radius diagrams constructed from the NL3* parametrization for different values of $\beta_3$ and $k_F^{\mbox{\tiny DM}}$, namely, (a) $k_F^{\mbox{\tiny DM}} = 0$ (no DM included), (b) $k_F^{\mbox{\tiny DM}} = 0.02$~GeV, and (c) $k_F^{\mbox{\tiny DM}} = 0.04$~GeV, and (d) $k_F^{\mbox{\tiny DM}} = 0.06$~GeV. The contours are the same exhibited in Fig~\ref{Fig1}.}
\label{Fig2}
\end{figure*}

In order to analyze how the neutron star properties behave with the increase of dark matter contribution to the system, an important step is to provide mass-radius diagrams with fixed values for the free parameter $\beta_{3}$, as depicted in Fig.~\ref{Fig1}. 
From this figure, it is possible to verify that the maximum neutron star mass decreases when more dark matter content is included in the system, for all $\beta_3$ values presented. This feature is observed even for the case of $\beta_3=1$ (Fig.~\ref{Fig1}{\color{blue}b}, General Relativity) and was reported already in recent literature, see Refs.~\cite{rmfdm2,rmfdm3,abdul,rmfdm6,rmfdm8,rmfdm10,rmfdm11} for instance. Consequently, the neutron stars' radii decrease with $k_F^{\mbox{\tiny DM}}$ and there will be a natural limit for this quantity to enable the agreement between the mass-radius profiles and recent observational astrophysical data (band regions also displayed in the figure). Furthermore, by analyzing the MTG studied here, i.e., $\beta_3\ne 1$, we notice that an increase of $\beta_3$ induces an increase in both, mass and radius for the neutron stars generated. This feature can be more easily identified in Fig.~\ref{Fig2}.
In this figure, we plot the mass-radius profiles by keeping fixed DM content in each panel. The case of no DM included in the system, $k_F^{\mbox{\tiny DM}} = 0$, is also shown in Fig.~\ref{Fig2}{\color{blue}a} for comparison. In summary, we verify that the decrease of the neutron stars' mass caused by the increase of the DM contribution is balanced by the increase of $\beta_3$. Therefore, it is possible for the system to support more DM in comparison with the General Relativity case. For the sake of completeness, we provide in Table~\ref{tab:beta3_ranges} the values obtained for $\beta_3$ that produced mass-radius diagrams fully compatible with each particular astrophysical constraint analyzed, namely, gravitational waves data related to the GW170817~\citep{Abbott_2017,Abbott_2018}, GW1908014~\citep{Abbott_2020}, and GW190425~\cite{Abbott_2020-2} events, some of them provided by the LIGO and Virgo Collaboration; data from the NASA's Neutron star Interior Composition Explorer (NICER) mission regarding the pulsars PSR~J0030+0451~\citep{Riley_2019,Miller_2019} and PSR~J0740+6620~\citep{Riley_2021,Miller_2021}; data from the latter pulsar extracted from Ref.~\citep{Fonseca_2021}; and data from PSR~J0952-0607~\cite{Romani_2022}.
\npdecimalsign{.}
\nprounddigits{2}
\begin{table*}[!htb]
\tabcolsep=0.30cm
\def\arraystretch{1.2}
\centering
\caption{Minimum and maximum neutron stars properties values of $\beta_3$ for different DM contributions.}
\begin{tabular}{l|cc|n{2}{2}n{2}{2}|n{2}{2}n{2}{2}|n{2}{2}n{2}{2}|n{2}{2}n{2}{2}}
\hline\hline
Events  & \multicolumn{2}{c|}{$\beta_3$} & \multicolumn{2}{c|}{$M_{\rm max}~($M$_\odot$)} & \multicolumn{2}{c|}{$R_{\rm max}$~(km)} & \multicolumn{2}{c|}{$R_{1.4}$~(km)} & \multicolumn{2}{c}{$R_{1.6}$~(km)}\\ 
\cline{2-11}          
          & min & max & min & max & min & max & min & max & min & max\\
\hline
\multicolumn{11}{c}{$k_{\rm F}^{\rm DM} = 0$} \\ 
\hline
GW190814               & 0.820 & 0.930 & 2.50359631 & 2.66623759 & 12.0172291 & 12.7888708 & 13.3525 & 14.2243 & 13.3423 & 14.2110\\
PSR J0952-0607         & 0.620 & 0.830 & 2.17697382 & 2.51881599 & 10.4485979 & 12.0859098 & 11.5989 & 13.4325 & 11.5595 & 13.4201\\
Lines PSR J0704+6620   & 0.530 & 0.610 & 2.01277542 & 2.15934634 & 9.65172768 & 10.3638020 & 10.7030 & 11.5007 & 10.6354 & 11.4581\\
Elipses PSR J0704+6620 & 0.540 & 1.940 & 2.03167510 & 3.85086465 & 9.74935341 & 18.4813080 & 10.808 & 20.6104 & 10.7480 & 20.5682\\
GW170817               & 0.360 & 0.830 & 1.65885580 & 2.51881599 & 7.95636225 & 12.0859098 & 8.717 & 13.4325 & 7.3428 & 13.4201\\
PSR J0030+0451         & 0.450 & 1.170 & 1.85465717 & 2.99054480 & 8.89420128 & 14.3538313 & 9.8325 & 15.9564 & 9.7129 & 15.9466\\
GW190425               & 0.560 & 1.100 & 2.06895661 & 2.89970446 & 9.92798615 & 13.9128246 & 11.0163 & 15.4718 & 10.9546 & 15.4677\\
\hline
\multicolumn{11}{c}{$k_{\rm F}^{\rm DM} = 0.02$~(GeV)} \\ 
\hline
GW190814               & 0.840 & 0.960 & 2.49781656 & 2.67027831 & 11.8806238 & 12.7037411 & 12.9880 & 13.8587 & 13.0254 & 13.9021\\
PSR J0952-0607         & 0.640 & 0.850 & 2.18027306 & 2.51264048 & 10.3661957 & 11.9541845 & 11.3701 & 13.0552 & 11.3751 & 13.1045\\
Lines PSR J0704+6620   & 0.540 & 0.620 & 2.00270867 & 2.14593601 & 9.53037167 & 10.2087727 & 10.4524 & 11.1946 & 10.4218 & 11.1941\\
Elipses PSR J0704+6620 & 0.560 & 2.110 & 2.03945875 & 3.95879102 & 9.70107079 & 18.8505650 & 10.6443 & 20.3245 & 10.6204 & 20.3879\\
GW170817               & 0.390 & 0.900 & 1.70197511 & 2.58548570 & 8.09228802 & 12.2949686 & 8.8383 & 13.4310 & 7.2224 & 13.4781\\
PSR J0030+0451         & 0.490 & 1.270 & 1.90773892 & 3.07130361 & 9.07198238 & 14.6153212 & 9.9559 & 15.8664 & 9.8934 & 15.9381\\
GW190425               & 0.590 & 1.190 & 2.09337449 & 2.97299647 & 9.95231247 & 14.1419039 & 10.9261 & 15.3778 & 10.9112 & 15.4337\\
\hline
\multicolumn{11}{c}{$k_{\rm F}^{\rm DM} = 0.04$~(GeV)} \\ 
\hline
GW190814               & 1.010 & 1.150 & 2.50870562 & 2.67693567 & 11.6610308 & 12.4410200 & 12.1719 & 12.9050 & 12.3284 & 13.0723\\
PSR J0952-0607         & 0.760 & 1.020 & 2.17618489 & 2.52109432 & 10.1185741 & 11.7144194 & 10.6960 & 12.2292 & 10.8054 & 12.3841\\
Lines PSR J0704+6620   & 0.650 & 0.750 & 2.01254535 & 2.16182041 & 9.35384083 & 10.0492163 & 9.9533 & 10.6358 & 10.0134 & 10.7365\\
Elipses PSR J0704+6620 & 0.700 & 2.940 & 2.08851695 & 4.28018618 & 9.71084118 & 19.9007397 & 10.3025 &  19.7409 & 10.3852 & 19.9734\\
GW170817               & 0.520 & 1.260 & 1.80007529 & 2.80203962 & 8.36620808 & 13.0260334 & 8.9536 & 13.4546 & 8.9291 & 13.6253\\
PSR J0030+0451         & 0.680 & 1.750 & 2.05846477 & 3.30223536 & 9.57201767 & 15.3552532 & 10.1649 & 15.5988 & 10.2395 & 15.8054\\
GW190425               & 0.770 & 1.640 & 2.19045520 & 3.19676661 & 10.1774836 & 14.8625278 & 10.7612 & 15.1503 & 10.8667 & 15.3445\\
\hline
\multicolumn{11}{c}{$k_{\rm F}^{\rm DM} = 0.06$~(GeV)} \\ 
\hline
GW190814               & 1.420 & 1.620 & 2.50162792 & 2.67199755 & 11.3580246 & 12.1238403 & 11.4172 & 12.0707 & 11.6343 & 12.3132\\
PSR J0952-0607         & 1.080 & 1.450 & 2.18167949 & 2.52791548 & 9.90790367 & 11.4754133 & 10.1538 & 11.5167 & 10.3285 & 11.7464\\
Lines PSR J0704+6620   & 0.920 & 1.050 & 2.01359892 & 2.15116501 & 9.14256859 & 9.76032639 & 9.4690 & 10.0302 & 9.6018 & 10.2012\\
Elipses PSR J0704+6620 & 1.040 & 4.860 & 2.14089680 & 4.62804079 & 9.71743011 & 21.0155964 & 9.9891 & 19.3389 & 10.1557 & 19.6844\\
GW170817               & 0.760 & 2.090 & 1.83014655 & 3.03495455 & 8.31087303 & 13.7746029 & 8.6954 & 13.4619 & 8.7608 & 13.7331\\
PSR J0030+0451         & 1.120 & 2.890 & 2.22171354 & 3.56884742 & 10.0869808 & 16.2046146 & 10.3145 & 15.4511 & 10.4914 & 15.7632\\
GW190425               & 1.190 & 2.700 & 2.29008985 & 3.44953799 & 10.4002209 & 15.6645765 & 10.5913 & 15.0063 & 10.7776 & 15.3124\\
\hline
\multicolumn{11}{c}{$k_{\rm F}^{\rm DM} = 0.08$~(GeV)} \\ 
\hline
GW190814               & 2.190 & 2.500 & 2.50054669 & 2.67167068 & 11.1639242 & 11.9200392 &  10.9668 & 11.5811 & 11.2282 & 11.8575\\ 
PSR J0952-0607         & 1.660 & 2.230 & 2.17704272 & 2.52327943 & 9.71061611 & 11.2650080 & 9.7836 & 11.0483 & 9.9881 & 11.3109\\
Lines PSR J0704+6620   & 1.410 & 1.620 & 2.00642300 & 2.15065336 & 8.95533466 & 9.60227871 & 9.1359 & 9.6852 & 9.2997 & 9.8879\\
Elipses PSR J0704+6620 & 1.690 & 8.420 & 2.19662666 & 4.90307999 & 9.80101395 & 21.8903275 & 9.8534 & 19.1368 & 10.0695 & 19.5481\\
GW170817               & 1.230 & 3.630 & 1.87398231 & 3.21933556 & 8.35858822 & 14.3675871 & 8.6139 & 13.4932 & 8.7449 & 13.8208\\
PSR J0030+0451         & 1.930 & 4.980 & 2.34742427 & 3.77074862 & 10.4772730 & 16.8335056 & 10.4136 & 15.3655 & 10.6509 & 15.7283\\
GW190425               & 1.960 & 4.680 & 2.36559796 & 3.65540791 & 10.5618105 & 16.3117752 & 10.4783 & 14.9709 & 10.7157 & 15.3339\\
\hline\hline
\end{tabular}
\label{tab:beta3_ranges}
\end{table*}

\section{Summary and concluding remarks}
\label{summ}

In this work, we have analyzed the effects of dark matter in neutron stars in the context of a modified teleparallel gravity theory. The dark particle considered here, namely, the lightest neutralino, is described by a fermion field that interacts with the Higgs boson. This mediator is also allowed to interact with ordinary matter, more specifically, with the nucleon Dirac fermion field. The equation of state of the complete fluid comprising dark matter, leptons, hadrons, and Higgs bosons is dependent on the dark matter Fermi momentum and was used to solve TOV-like equations in the context of a modified theory of gravity. The latter, by its turn, is an extension of the TEGR theory built with a linear combination with arbitrary coefficients ($\beta_{1},\beta_{2},\beta_{3}$) of the three quadratic invariants that compose the TEGR Lagrangian. In order to be compatible with the covariant conservation of the energy-momentum tensor, a constraint is imposed on these three arbitrary coefficients. This leaves us with only one parameter that can be interpreted as a modulation of the gravitational coupling constant.

With the set of teleparallel field equations and the input EoS, several mass-radius diagrams were built for different values of the parameters~$k_F^{\mbox{\tiny DM}}$ (Fermi momentum of the dark particle) and~$\beta_{3}$. These curves were plotted against the contours related to data obtained from several observations/measurements (NICER, PSR~J0030+0451, PSR~J0740+6620, PSR~J0920-0607, GW170817, GW190814, and GW190425), which allowed us to establish limits for $\beta_{3}$ for different values of $k_F^{\mbox{\tiny DM}}$.
We verified that the maximum neutron star mass decreases with $k_F^{\mbox{\tiny DM}}$, while it increases with $\beta_{3}$. Essentially, we conclude that $\beta_{3}>1$ allows us to accommodate more dark matter content in the neutron star in comparison with the case of GR ($\beta_3=1$).

As a next step, we intend to use Bayesian inference in order to establish a confidence region in the parameter space $k_F^{\mbox{\tiny DM}}$ vs $\beta_{3}$ which allows us to accommodate the available astrophysical data.

\section*{ACKNOWLEDGMENTS}
This work is a part of the project INCT-FNA proc. No. 464898/2014-5. It is also supported by Conselho Nacional de Desenvolvimento Cient\'ifico e Tecnol\'ogico (CNPq) under Grants No. 312410/2020-4 (O.~L.), No. 308528/2021-2 (M.~D.). S.~G.~V and P.~J.~P. are grateful to CNPq for financial support under Grant No. 400879/2019-0.. O.~L., and M.~D. also acknowledge Funda\c{c}\~ao de Amparo \`a Pesquisa do Estado de S\~ao Paulo (FAPESP) under Thematic Project 2017/05660-0. O.~L. is also supported by FAPESP under Grant No. 2022/03575-3 (BPE). This study was financed in part by the Coordena\c{c}\~ao de Aperfei\c{c}oamento de Pessoal de N\'ivel Superior - Brazil (CAPES) - Finance Code 001 - Project number 88887.687718/2022-00 (M.~D.).

\bibliographystyle{apsrev4-2}
\bibliography{references}

\end{document}